\documentclass{article}
\usepackage{spconf,amsmath,epsfig}
\usepackage{url}

\usepackage{mathtools}

\usepackage{textcomp}
\usepackage{microtype}
\usepackage{booktabs,makecell,tabularx}

\newcolumntype{C}[1]{>{\centering\arraybackslash}p{#1}}
\newcolumntype{L}{>{\raggedright\arraybackslash}X}

\usepackage{siunitx}
\usepackage{etoolbox}
\usepackage{multirow}
\usepackage[export]{adjustbox}
\usepackage{flushend}
\newrobustcmd{\B}{\bfseries}

\title{On the Modelling of Ship Wakes in S-Band SAR Images\\and an Application to Ship Identification}
\name{Kamirul Kamirul\sthanks{The authors would like to thank SSTL and Airbus D\&S for the provision of NovaSAR-1 data. Contact author e-mail address: \texttt{kamirul.kamirul@bristol.ac.uk}}, Odysseas Pappas, Igor G. Rizaev, Alin Achim }
\address{Visual Information Laboratory\\	University of Bristol, UK}

\begin{document}
%
\maketitle
\begin{abstract}
We present a novel ship wake simulation system for generating S-band Synthetic Aperture Radar (SAR) images, and demonstrate the use of such imagery for the classification of ships based on their wake signatures via a deep learning approach. Ship wakes are modeled through the linear superposition of wind-induced sea elevation and the Kelvin wakes model of a moving ship. Our SAR imaging simulation takes into account frequency-dependent radar parameters, i.e., the complex dielectric constant ($\varepsilon$) and the relaxation rate ($\mu$) of seawater. The former was determined through the Debye model while the latter was estimated for S-band SAR based on pre-existing values for the L, C, and X-bands. The results show good agreement between simulated and real imagery upon visual inspection. The results of implementing different training strategies are also reported, showcasing a notable improvement in accuracy of classifier achieved by integrating real and simulated SAR images during the training.
\end{abstract}
\begin{keywords}
Ship wakes, sea waves, SAR simulation, S-band, NovaSAR-1, vessel classification, sea modelling
\end{keywords}
\section{Introduction}
\label{sec:INTRODUCTION}
Synthetic Aperture Radar (SAR) technology exhibits weather-independent performance and  enables observations in any atmospheric and illumination conditions~\cite{rs11121456}. SAR imaging is widely used in a variety of applications, including maritime surveillance, environmental monitoring, and disaster response~\cite{rs12142190,7131452}. 

Ship wake detection in SAR images is gaining popularity as the ship wake signatures can serve as a way to estimate ship dynamics parameters and can also be utilized for more practical applications, e.g., building a model for ship classification based on the wakes~\cite{9323097,HEISELBERG2023113492}. This latter application unlocks the capability to identify ship classes from SAR images, offering the potential for advanced maritime surveillance capabilities. However, developing such classification models requires a huge volume of training images, which can be challenging to obtain due to the limited availability of publicly accessible datasets. 

The majority of SAR wake simulation literature has primarily focused on L-, C-, and X-band radar, as these are used by some of the most widely utilised SAR satellite missions (such as TerraSAR-X, COSMO-SkyMed, ICEYE, ALOS-2, ERS-2, and SEASAT~\cite{RIZAEV2022120,Tunaley_1991, Nunziata, Hennings_Romeiser_Alpers}). However, new satellite missions such as NovaSAR-1 and NISAR-S utilise S-band instruments due to their advantages for maritime monitoring applications. Motivated by the above, our work focuses on simulating ship wakes specifically as imaged under a S-Band SAR system, and then utilizes the resulted images for classification purposes in real scenarios. Therefore, the simulation results from this work will not only supplement the existing modelling in terms of the frequency band used but also provide insights into the characteristics of ship wake signatures in the S-band scenario.

Ship wake simulation under S-Band SAR system have not been previously conducted largely due to limited information on the frequency-dependent parameters used for the simulation, and the absence of real datasets for verifying simulation results. Two essential frequency-dependent parameters, i.e., the complex dielectric constant ($\varepsilon$) and relaxation rate ($\mu$) for S-Band were estimated and provided in this work. These parameters were then used for calculating the normalized radar cross-section (NRCS) of sea surface and hydrodynamic modulation effect respectively. The complex dielectric constant $\varepsilon$ for S-band SAR was obtained using the Debye approach, as reported by Meissner and Wentz in ~\cite{Meissner_2000, Meissner_2004}, which is based on an extensive experimental dataset for seawater. Meanwhile, the value of $\mu$ was estimated from pre-existing values in the L, C, and X-bands, according to ~\cite{RIZAEV2022120} and ~\cite{Bruning_1994}.  

To demonstrate the utility of the simulated SAR images, we employed the images to train a deep-learning model, AlexNet, for classifying real ships in SAR images acquired by NovaSAR-1. Different training strategies, including transfer learning and mixing of simulated and limited numbers of real images in the training dataset, have been conducted and evaluated concerning their impact on classification performance.

In summary, our contributions are two-fold. Firstly, to the best of our knowledge, we are the first to conduct ship wakes simulation for S-Band SAR. Therefore, our simulation can serve as a theoretical foundation for future work. Secondly, the training strategies outlined in our work provide a pathway for utilizing simulated ship wake images to build a model for classifying real ship wakes images.To demonstrate the utility of simulated SAR images, we employed the images to train a deep-learning model, AlexNet, for classifying real ship images acquired by NovaSAR-1.  
 
\section{MATERIALS AND METHODS}
\label{sec:MATERIALS_AND_METHODS}

This work employs prior work on ship wake generation, which we have here extended to S-band SAR; a comprehensive explanation of the previous simulator can be found in ~\cite{RIZAEV2022120} while code can be accessed through the University of Bristol Research Data Repository at \url{doi.org/10.5523/bris.el0p94vgxjhi2224bx78actb4}.

\subsection{Sea Elevation Model}
\label{ssec:SeaElevationModel}

In this work, the sea elevation model is formulated as the superposition of the wind-induced sea surface and the Kelvin wakes generated by a moving ship. Wind-induced sea elevation at time \textit{t} can be expressed as~\cite{RIZAEV2022120,Bruning_1994,9327469}:
\begin{flalign}
Z_{sea} & =\sum_{i}\sum_{j} A_{ij}\nonumber&\\&\cos\left({k_{i}\left[ 
x\cos{\theta_{i}} + y\sin{\theta_{j}}+\omega t + \sigma_{ij}\right]}  \right),
\end{flalign}where \textit{k} and $\omega$ respectively are the wavenumber and radial frequency while $\sigma$ is the initial phase uniformly distributed between (0, 2$\pi$). $A_{ij}$ represents the amplitude expressed as:
\begin{equation}
A_{ij}=\sqrt{2S(k_{i})D(k_{i}\theta_{j})\Delta{k_{i}\Delta{\theta_{j}}}},
\end{equation}where $S(k)$ and $D(k_{i}\theta_{j})$ are wave spectrum and angular spreading function~\cite{RIZAEV2022120}. 

In modelling the waves generated by moving ships, we used the Kelvin waves model,$Z_{ship}$, represented by Zilman et al. in ~\cite{Zilman}: 

\begin{equation}
Z_{ship}=\frac{V_{s}}{g}\frac{\partial{\Phi_{s}}}{\partial{x}},
\end{equation}with $V_{s}$ and $\Phi_{s}$ being ship velocity and an approximated form of fluid velocity potential. Finally, composite sea elevation model ($Z$) due to existing wind and moving ship can be obtained as the superposition of $Z_{sea}$ and $Z_{ship}$.

\subsection{Frequency-Dependent Parameters}
\label{ssec:FrequencyDependentParameters}
There are two essential parameters that need to be determined to model the interaction between incoming waves and the sea surface, i.e., complex dielectric constant ($\varepsilon$) and the relaxation rate ($\mu$).

The complex dielectric constant ($\varepsilon$) has been calculated based on the work of Meissner and Wentz in ~\cite{Meissner_2000, Meissner_2004}. Therein, extensive experiments were conducted to determine $\varepsilon$ for both pure and sea water over a temperature range from -20°C to +40°C and a frequency range up to 90 GHz. The dielectric constant ($\varepsilon$) of sea water  with a specific salinity ($S$) and temperature ($T$), radiated under a specific wavelength ($\lambda$), can be calculated as:
\begin{flalign}
\varepsilon(\lambda,S,T)&=\nonumber&\\&\varepsilon_{R} + \frac{\varepsilon_{S}(S,T)-\varepsilon_{R}}{1+{\left[i\lambda_{R}(S,T)/\lambda\right]}^{1-\alpha}}-2\frac{i\sigma(S,T)\lambda}{c},
\end{flalign}Here, $\lambda_{R}$, $\varepsilon_{S}$, and $\sigma$ are all dependent on $S$ and $T$, while $\varepsilon_{R}$ and $\alpha$ are set to 4.44 and 0.012, respectively, and $c$ represents the speed of light. The detailed implementation of above equation can be found on ~\cite{Meissner_2000}. Assuming that $S = 35$ ppt~\cite{Global_Salinity}, $T=21^{\circ}C$~\cite{Global_SST}, and NovaSAR-1 frequency is 3.2 GHz~\cite{NovaSAR_Specifications_01}, the corresponding dielectric constant for our simulation is 69.63-\textit{i}38.95. 

The magnitude of $\mu$ is poorly known experimentally, especially for S-Band case. In our experiment, the $\mu$ value for S-band was interpolated from the known $\mu$ values for L, C, and X-band, which were provided in ~\cite{Bruning_1994,Caponi_1988}. Finally, the estimated value for $\mu$ in the NovaSAR-1 band are $\mu=0.05$ s\textsuperscript{-1} for $V_w$ $\le$ 5 m/s and $\mu=0.39$ s \textsuperscript{-1} for  $V_w$ $>$ 5 m/s.

\subsection{SAR Imaging of Sea Surface}
\label{ssec:SARImagingofSeaSurface}

To simulate SAR images of the modeled sea surface, we employ the two-scale model (TSM). This model is based on the resonant Bragg scattering theory describing the scattering of electromagnetic waves from a rough surface~\cite{RIZAEV2022120,9327469}. The Bragg scattering solution for normalized radar cross-section (NRCS) with VV and HH polarizations is expressed as ~\cite{RIZAEV2022120}:
\begin{equation}
    \sigma_{0}=8\pi k_e^4 \cos^4{\theta_{l}}W(k_{Bx},k_{By})|T_c^2|,
\end{equation}where $k_e$ is radar wavenumber. The $\theta_l$, $W$, and $T_c$ are respectively local incidence angle, wavenumber spectral density and complex scattering function. Then, the NRCS after taking into account hydrodynamic and tilt modulations:
\begin{equation}
\label{aaa}
    \sigma(x,y)=\sigma_{0}\left[ 1 + \int{(M(k)F(k)e^{ikx}+c.c)dk} \right],
\end{equation}where $F(k)$ is the 2D FFT of composite sea elevation model $Z$. $M(k)$ is defined as follow  ~\cite{Bruning_1994,Lyzenga01}:
\begin{equation}
    M(k)=-4.5\frac{k_y^2}{|k|}\frac{\omega-i\mu}{\omega+i\mu}+\frac{4\cot{\theta_r}}{1\pm \sin^2{\theta_r}}ik_{y},
\end{equation}where $\theta_r$ is the nominal radar incident angle. The SAR image intensity can then be obtained using:
\begin{flalign}
I(x,y) & =\iint{\partial(y_i-y)}\nonumber&\\&\frac{\sigma(x,y)}{P_a^{'}(x,y)}{\exp{\left(-\pi^2 {\left[ {\frac{x_i-x-\frac{R}{V}U_r}{b}}\right]}^2\right)}} dxdy.
\end{flalign}

$\partial(.)$ is the Dirac delta function of the impulse response function in range direction while $R$, $H$, and $V$ are the range distance, platform height, and platform velocity, respectively. $P_a^{'}$ is degraded azimuth resolution which depends on number of incoherent looks and radar wavelength ~\cite{RIZAEV2022120}. Finally, multiplicative random noise is added to the final image to simulate speckle, using the Weibull probability density function (PDF) as described in ~\cite{Karakus}.

\subsection{Ship Classification}
\label{ssec:ShipClassification}
In this work, the AlexNet \cite{AlexNet} model was trained to differentiate wakes produced by cargo and tanker ships. Due to the limited availability of real ship wake images for testing, we utilized ship wake crops from five real NovaSAR-1 products and applied data augmentation by rotating, flipping, and varying noise amplitude. In total, this dataset consisted of 960 real images and  is hence denoted as \textbf{NVReal} dataset.

Three different strategies for training the model have been conducted. Both \textbf{Experiment I} and \textbf{Experiment II} are based on a transfer learning approach. For this purpose, a pre-trained model was produced using a large dataset, \textbf{SynthwakeSAR}, a synthetic dataset contains a total of 46,080 simulated SAR ship wakes images  ~\cite{SynthwakeSAR_Paper}. \textbf{SynthwakeSAR} dataset is accessible at \url{doi.org/10.5523/bris.30kvuvmatwzij2mz1573zqumfx}. Since the \textbf{SynthwakeSAR} dataset was originally generated for X-band SAR, we reproduced the dataset to represent images acquired by an S-band SAR system (denoted as \textbf{SynthwakeSAR-S} dataset in this work) and used that for initial training. Afterwards, a simulated NovaSAR-1 dataset (\textbf{NVSim}) was generated based on parameters of real ship wake images from NovaSAR-1 (\textbf{NVReal}). This \textbf{NVSim} dataset was  used to re-train the pre-trained model via fine-tuning (\textit{FT}) for \textbf{Experiment I} and via feature extraction (\textit{FE}) schemas for \textbf{Experiment II}. 

\textbf{NVSim} contains 18,000 images which were produced by assigning parameters within the range of the real parameters for the real NovaSAR images in \textbf{NVReal}, as shown in the Table \ref{table:NVSim_Parameters} (denoted as \textbf{R} superscripts). This step was to ensure that the simulated images closely resemble their real counterparts so that the classification model can learn their representation for recognizing the real images. Individual ship parameters are provided in Table \ref{table:Ships_Parameters}.

\begin{table}[!htb]
    \caption{Sea state and imaging parameters for generating \textbf{NVSim}.}
    \label{table:NVSim_Parameters}
    \sisetup{table-format=5.4}
    \centering
    \small
    \begin{tabular} {l S c c} 
        \toprule
\thead {Parameters}  
    & {\thead{Symbol} } 
        & {\thead{Value \\ range}} 
         & {\thead{Number of \\ sampling}} \\
    \midrule
    
Wind speed  10m & $U_{10}$ &  $U_{10}^{R}\pm0.5$ m/s &3\\
Wind direction  & $\theta_{w}$ &  $\theta_{w}^{R}\pm15^\circ$&10 \\
Ship speed      & $V_{s}$ & $V_{s}^{R}\pm0.5$ m/s&5 \\
Ship direction  & $\theta_{s}$ &  $\theta_{s}^{R}\pm30^\circ$&8 \\
Incident angle  & $\theta_{i}$ &  $\theta_{i}^{R}\pm2^\circ$ &3 \\
    \bottomrule
    \end{tabular}
\end{table}

\begin{table*}[!htb]
    \caption{Ship parameters for generating \textbf{NVSim}.}
    \label{table:Ships_Parameters}
    \sisetup{table-format=2.4}
    \centering
    \small
    \begin{tabular} {l c c c c c c c c c c} 
        \toprule
\thead {MMSI}  
    & {\thead{Class} }
    & {\thead{$L(m)$} } 
    & {\thead{$B(m)$}} 
    & {\thead{$T(m)$}}
    & {\thead{$Fetch(km)$}}
    & {\thead{$U_{10}^{R}$ (m/s)} } 
    & {\thead{$\theta_{w}^{R}$($^\circ$)}} 
    & {\thead{$V_{s}^{R}$ (m/s)}} 
    & {\thead{$\theta_{s}^{R}$($^\circ$)}} 
    & {\thead{ $\theta_{i}^{R}$($^\circ$)}}  
    \\
    \midrule
215071000    & Tanker & 183 & 32 & 9.30 & 80 &4.13 & 343.50 &6.73 & 351.50& 29.35\\
218433000    & Cargo & 126 & 23 & 6.90 & 25 &6.10 & 359.50 &7.45 & 182.50& 29.31\\
371720000    & Tanker & 229 & 103 &	9.50 & 80 &3.33 & 332.50 &7.56 & 355.50& 23.96\\
636013817    & Tanker & 227 & 150 & 9.70 & 80 &4.85 & 4.85 &6.89 & 156.50& 27.87\\
477665200    & Cargo & 266	& 150 & 10.50 & 80 &6.20 & 341.50 &8.44 & 181.50& 28.23\\
    \bottomrule
    \end{tabular}
\end{table*}

\begin{figure}
\includegraphics[width=.47\textwidth,center]{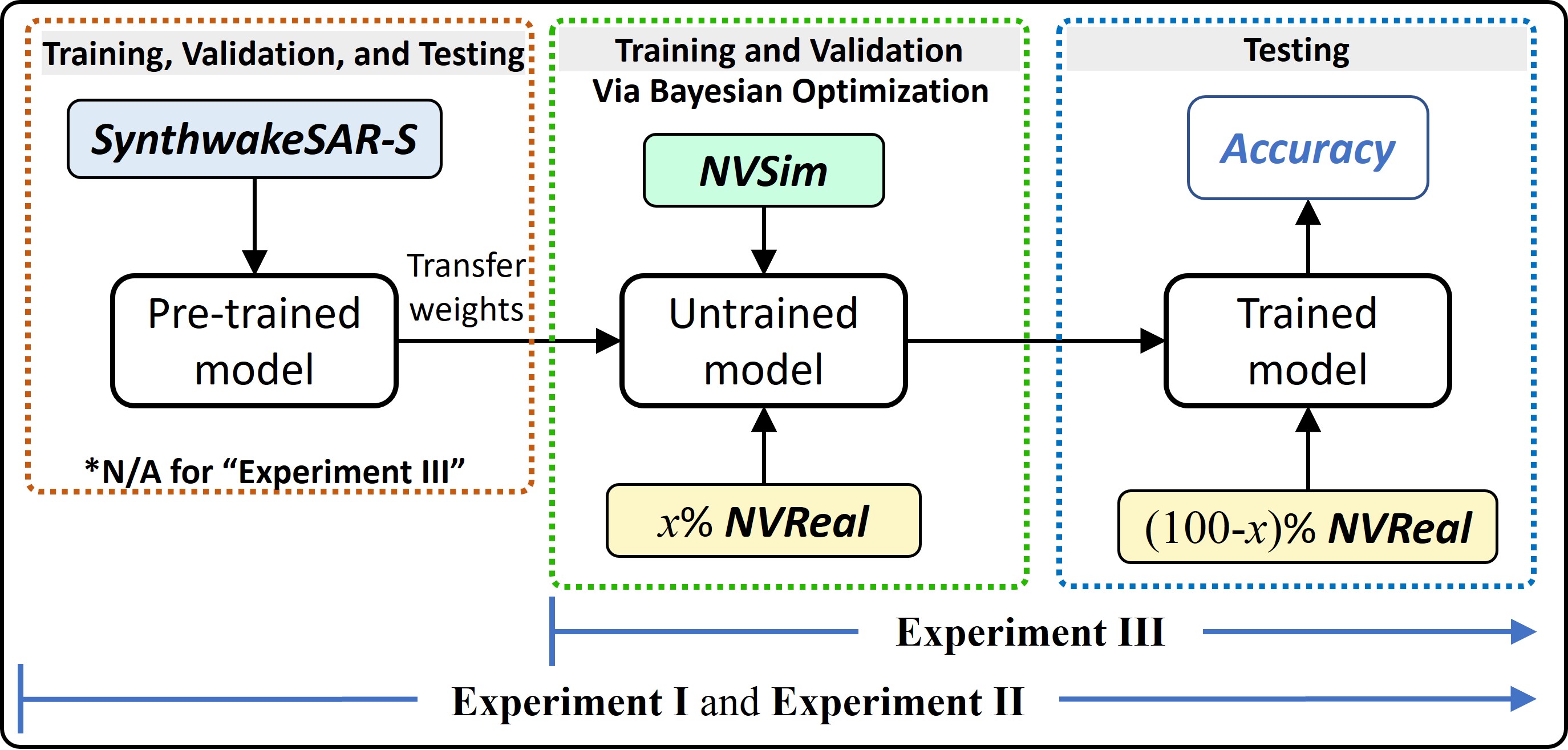}
\caption{Flowchart of experiments. The \textbf{SynthwakeSAR-S} refers to reproduced version of \textbf{SynthwakeSAR} using NovaSAR-1 imaging properties while \textbf{NVSim} and \textbf{NVReal} are respectively simulated and real datasets of ship wakes.}

\label{fig:Experiment_Flowchart}
\end{figure}

In contrast to \textbf{Experiment I} and \textbf{Experiment II}, \textbf{Experiment III} did not employ any pre-trained network. In this case, the model was directly trained using the \textbf{NVSim} dataset. This experiment aims to assess the model's performance in the absence of a pre-existing dataset. Furthermore, we incorporated a portion of images from \textbf{NVReal} and mixed them with \textbf{NVSim} during the training stage for all experiments. The amount of real images involved was varied from 10\% to 30\%. We did not extend beyond this range as it would reduce the number of the remaining \textbf{NVReal} for testing the model. The flowchart of the experiments is given in Figure \ref{fig:Experiment_Flowchart}.

\section{RESULTS AND DISCUSSION}
\label{sec:RESULTS_AND_DISCUSSION}

In this section, we present the visual comparison between the real and simulated images. Moreover, the performance of the classification model trained under different strategies is also discussed.

\subsection{Simulated NovaSAR-1 Images}
\label{ssec:SimulatedNovaSAR1Images}
In general, evaluating the accuracy of simulated images can be achieved through both qualitative and quantitative approaches. However, for our case, like-for-like comparison between simulated and real images is challenging because simulated images may not be entirely identical but rather akin or resembling the real ones. Therefore, measuring similarity between the images, for example using Structural Similarity (SSIM) index or correlation coefficient, would be inappropriate as these metrics work on a pixel-by-pixel basis. 

\begin{figure}[htb]
\includegraphics[width=.43\textwidth,center]{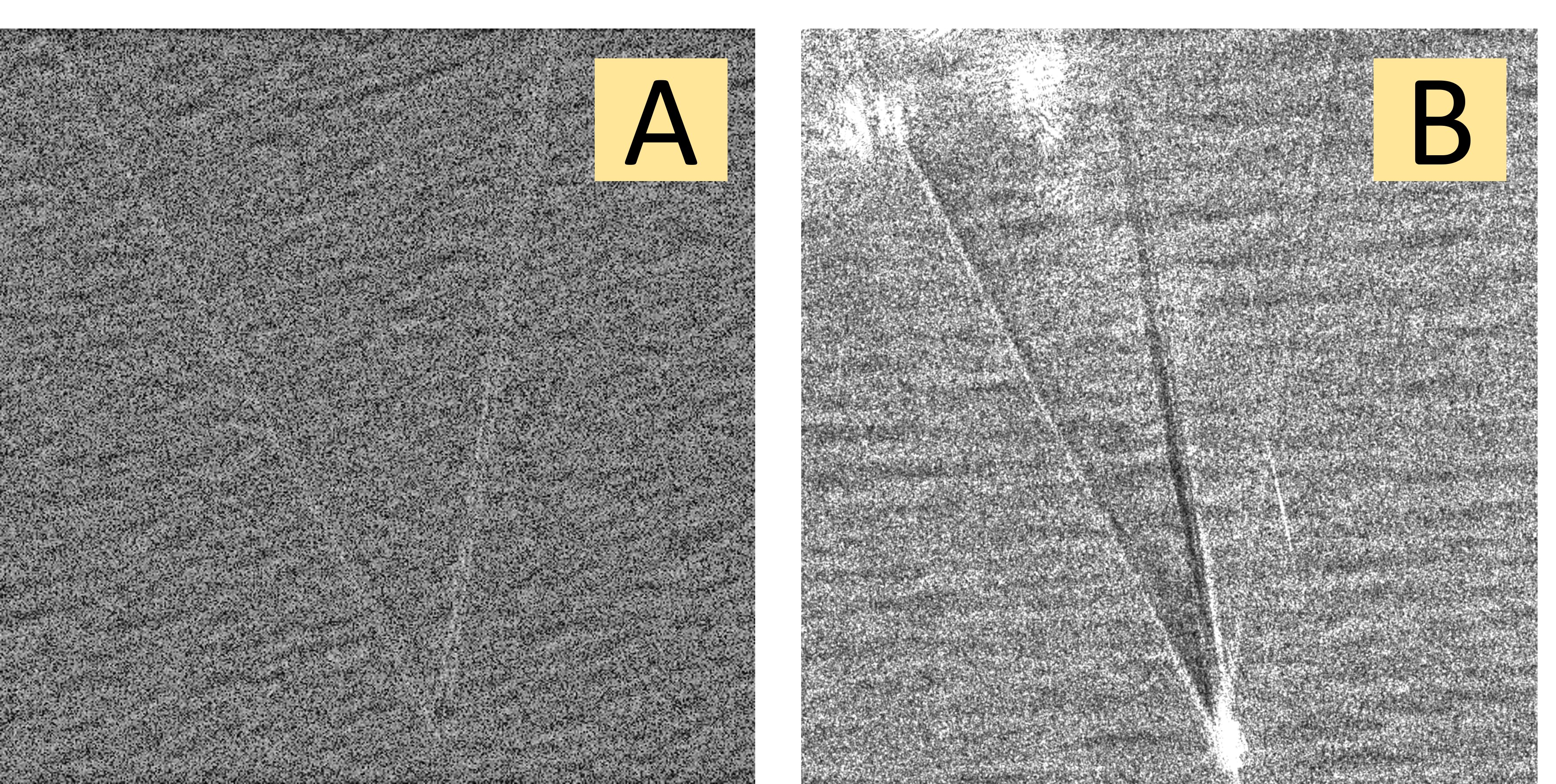}
\caption{Comparison between simulated (A) and real (B) images of ship wakes for NovaSAR-1 SAR system. The parameters used are: $L$=126 m, $B$=23 m, $T$=6.9 m, $\theta_{s}=182^{\circ}$, $V_{s}=7.45$ m/s, $\theta_{wind}=359.5^{\circ}$, $V_{w}= 6.10$ m/s, and $\theta_{i}=29.31^{\circ}$.}

\label{fig:Real_vs_Sim}
\end{figure}

Considering the aforementioned limitation, we chose to compare the images through visual inspection. As depicted in Figure \ref{fig:Real_vs_Sim}, the simulated image closely resembles the real one, except for the appearance of turbulent wake on the real images. Furthermore, as the ship moves faster and parallel to the platform direction, the height of wind-induced ambient waves increases, resulting in the transverse waves becoming scarcely visible. This phenomenon leaves only the prominent Kelvin arms visible in both images, indicating a consistent pattern or agreement between the two images. This behavior has also been noted in ~\cite{RIZAEV2022120,Tings_001}. 

\subsection{Classification Results and Performance}
\label{ssec:ClassificationResultsandPerformance}
In this section, the use of simulated images for classifying real images is demonstrated. Table \ref{table:Classification_Performance} shows the classification accuracy for each experimental setup under different proportions of \textbf{NVReal} included for training the model. 

\begin{table}[!htb]
    \caption{Classification performance for all three experiments.}
    \label{table:Classification_Performance}
    \sisetup{table-format=5.4}
    \centering
    \small
    \begin{tabular} {c c c c} 
        \toprule
\thead {\multirow{2}{*}{NVReal Portion (\%)}}  
        & \multicolumn{3}{c}{Accuracy (\%)} \\ 
        \cmidrule{2-4}
     & \textbf{Exp. I} & \textbf{Exp. II} & \textbf{Exp. III} \\
         
    \midrule

0  & 60.41 & 50.86 & 51.56 \\
10 & 60.25 & 60.41 & 63.96 \\
20 & 40.02 & 64.14 & 59.97 \\
30 & 68.55 & 72.13 & 71.83 \\
    \bottomrule
    \end{tabular}
\end{table}

Based on the experimental results, three key findings have been highlighted. Firstly, the model accuracy is generally low if no real images are introduced to the model during training. While it may be assumed that this result suggests a poor representation of real images in the simulated dataset, it is noteworthy that the accuracy across all experiments remains above 50\%. This implies that training the model solely using \textbf{NVSim} does provide actionable information and is not reduced to random guessing. 

Secondly, introducing a portion of real images in the training process can significantly boost the model performance. It is found that adding 30\% of the available \textbf{NVReal} data to \textbf{NVSim} in the training process improves the accuracy by almost 12\% for \textbf{Experiment II}. This result indicates that it is still possible to build a good classification model even when there is limited real data available for training. 

Lastly, fine-tuning a model pre-trained under synthetic ship wake dataset (i.e, \textbf{Experiment II}) is capable of producing a higher accuracy of up to 72\%. This result suggests that combining representative simulated dataset and re-training all the network weights could provide benefits for building a classification model in the absence of a real dataset for training. 

It should be noted that our results are based on experiments conducted on five ship models due to the challenges in obtaining real images along with their ship and sea states, as well as imaging parameters. Therefore, a more refined result is likely to be obtained with an increased number of real datasets. This would enhance the model's ability to learn features from a wider range of real images rather than relying extensively on augmented images. Nevertheless, our results are presented to showcase the benefits of using simulated images in developing a deep learning model for classifying real ships in SAR images.

\section{CONCLUSIONS}
\label{sec:CONCLUSIONS}
In this work, the first results on simulation of ship wake acquired under S-Band SAR system is reported. The simulation has been conducted by employing a two-scale model, which requires frequency-dependent parameters, i.e., the complex dielectric constant ($\varepsilon$) and the relaxation rate ($\mu$) of seawater. The value of $\varepsilon$ was determined using the Debye model, whereas $\mu$ was estimated for the S-band based on pre-existing values in the L, C, and X-bands. 

Based on visual comparison, there is good agreement between the simulated and real images. Additionally, we showcased the utilization of simulated images to address a real-world challenge, i.e., ship wake classification on NovaSAR-1 images. We have found that a classification model having accuracy of up to 72\% can be attained by incorporating a very small proportion of real images along with simulated ones in the training stage.

\bibliographystyle{IEEEbib}
\bibliography{refs}

\begin{thebibliography}{10}

\bibitem{rs11121456}
Y.~S. Tsai, A.~Dietz, N.~Oppelt, and C.~Kuenzer,
\newblock ``{Remote Sensing of Snow Cover Using Spaceborne SAR: A Review},''
\newblock {\em Remote Sensing}, vol. 11, no. 12, 2019.

\bibitem{rs12142190}
S.~Adeli, B.~Salehi, M.~Mahdianpari, L.~J. Quackenbush, B.~Brisco, H.~Tamiminia, and S.~Shaw,
\newblock ``{Wetland Monitoring Using SAR Data: A Meta-Analysis and Comprehensive Review},''
\newblock {\em Remote Sensing}, vol. 12, no. 14, 2020.

\bibitem{7131452}
E.~Makhoul, S.~V. Baumgartner, M.~Jäger, and A.~Broquetas,
\newblock ``{Multichannel SAR-GMTI in Maritime Scenarios With F-SAR and TerraSAR-X Sensors},''
\newblock {\em {IEEE J. Sel. Top. Appl. Earth Obs. Remote Sens.}}, vol. 8, no. 11, pp. 5052--5067, 2015.

\bibitem{9323097}
B.~Tings, S.~Wiehle, and S.~Jacobsen,
\newblock ``{Ship Wake Component Detectability on Synthetic Aperture Radar (SAR)},''
\newblock in {\em IGARSS 2020 - 2020 IEEE Int. Geosci. and Remote Sens. Symp.}, 2020, pp. 1233--1235.

\bibitem{HEISELBERG2023113492}
P.~Heiselberg, K.~Sørensen, and H.~Heiselberg,
\newblock ``{Ship Velocity Estimation in SAR Images Using Multitask Deep Learning},''
\newblock {\em Remote Sens. Environ.}, vol. 288, pp. 113492, 2023.

\bibitem{RIZAEV2022120}
I.~G. Rizaev, O.~Karakuş, S.~J. Hogan, and A.~Achim,
\newblock ``{Modeling and SAR Imaging of the Sea Surface: A Review of the State-of-the-art with Simulations},''
\newblock {\em ISPRS J. Photogramm. Remote Sens.}, vol. 187, pp. 120--140, 2022.

\bibitem{Tunaley_1991}
J.K.E. Tunaley, E.H. Buller, K.H. Wu, and M.T. Rey,
\newblock ``{The Simulation of the SAR Image of a Ship Wake},''
\newblock {\em IEEE Trans. Geosci. Remote Sens.}, vol. 29, no. 1, pp. 149--156, 1991.

\bibitem{Nunziata}
{F. Nunziata, A. Gambardella, and M. Migliaccio},
\newblock ``{An Educational SAR Sea Surface Waves Simulator},''
\newblock {\em Int. J. Remote Sens.}, vol. 29, no. 11, pp. 3051--3066, 2008.

\bibitem{Hennings_Romeiser_Alpers}
{I. Hennings, R. Romeiser, W. Alpers and A. Viola},
\newblock ``{Radar Imaging of Kelvin Arms of Ship Wakes},''
\newblock {\em Int. J. Remote Sens.}, vol. 20, no. 13, pp. 2519--2543, 1999.

\bibitem{Meissner_2000}
T.~Meissner and F.J. Wentz,
\newblock ``{Algorithm Theoretical Basis Document (ATBD) Version 2 :AMSR Ocean Algorithm },''
\newblock {\em EOS Project Goddard Space Flight Center, NASA}, pp. 1--3, 2000.

\bibitem{Meissner_2004}
T.~Meissner and F.J. Wentz,
\newblock ``{The Complex Dielectric Constant of Pure and Sea Water from Microwave Satellite Observations},''
\newblock {\em IEEE Trans. Geosci. Remote Sens.}, vol. 42, no. 9, pp. 1836--1849, 2004.

\bibitem{Bruning_1994}
C.~{Br{\"u}ning}, R.~{Schmidt}, and W.~{Alpers},
\newblock ``{Estimation of the Ocean Wave-radar Modulation Transfer Function from Synthetic Aperture Radar Imagery},''
\newblock {\em J. Geophys. Res.}, vol. 99, no. C5, pp. 9803--9815, May 1994.

\bibitem{9327469}
W.~Ren, P.~Liu, X.~Ren, and Y.~Jin,
\newblock ``{SAR Image Simulation of Ship Turbulent Wake Using Semi-Empirical Energy Spectrum},''
\newblock {\em IEEE J. Multiscale Multiphys. Comput. Tech.}, vol. 6, pp. 1--7, 2021.

\bibitem{Zilman}
G.~Zilman, A.~Zapolski, and M.~Marom,
\newblock ``{On Detectability of a Ship's Kelvin Wake in Simulated SAR Images of Rough Sea Surface},''
\newblock {\em IEEE Trans. Geosci. Remote Sens.}, vol. 53, no. 2, pp. 609--619, 2015.

\bibitem{Global_Salinity}
W.~J. Gould and S.~A. Cunningham,
\newblock ``{Global-scale Patterns of Observed Sea Surface Salinity Intensified Since the 1870s},''
\newblock {\em Commun. Earth Environ.}, vol. 2, no. 1, pp. 76, Apr 2021.

\bibitem{Global_SST}
``{Global Sea Surface Temperature Reaches a Record High},'' \url{https://climate.copernicus.eu/global-sea-surface-temperature-reaches-record-high}, 2023,
\newblock [Online; accessed 25-December-2023].

\bibitem{NovaSAR_Specifications_01}
R.~Bird, P.~Whittaker, B.~Stern, N.~Angli, M.~Cohen, and R.~Guida,
\newblock ``{NovaSAR-S: A Low Cost Approach to SAR Applications},''
\newblock in {\em Proc. of 2013 Asia-Pacific Conf. on Synthetic Aperture Radar (APSAR)}, 2013, pp. 84--87.

\bibitem{Caponi_1988}
E.~A. {Caponi}, D.~R. {Crawford}, H.~C. {Yuen}, and P.~G. {Saffman},
\newblock ``{{Modulation of Radar Backscatter from the Ocean by a Variable Surface Current}},''
\newblock {\em J. Geophys. Res.}, vol. 93, no. C10, pp. 12,249--12,263, Oct 1988.

\bibitem{Lyzenga01}
D.~R. Lyzenga,
\newblock ``{Numerical Simulation of Synthetic Aperture Radar Image Spectra for Ocean Waves},''
\newblock {\em IEEE Trans. Geosci. Remote Sens.}, vol. GE-24, no. 6, pp. 863--872, 1986.

\bibitem{Karakus}
O.~Karakuş, E.~E. Kuruoğlu, and A.~Achim,
\newblock ``{A Generalized Gaussian Extension to the Rician Distribution for SAR Image Modeling},''
\newblock {\em IEEE Trans. Geosci. Remote Sens.}, vol. 60, pp. 1--15, 2022.

\bibitem{AlexNet}
A.~Krizhevsky, I.~Sutskever, and G.~E. Hinton,
\newblock ``{ImageNet Classification with Deep Convolutional Neural Networks},''
\newblock in {\em Adv. Neural Inf. Process. Syst.} 2012, vol.~25, pp. 1097--1105, Curran Associates, Inc.

\bibitem{SynthwakeSAR_Paper}
I.~G. Rizaev and A.~Achim,
\newblock ``{SynthWakeSAR: A Synthetic SAR Dataset for Deep Learning Classification of Ships at Sea},''
\newblock {\em Remote Sensing}, vol. 14, no. 16, 2022.

\bibitem{Tings_001}
B.~Tings,
\newblock ``{Non-Linear Modeling of Detectability of Ship Wake Components in Dependency to Influencing Parameters Using Spaceborne X-Band SAR},''
\newblock {\em Remote Sensing}, vol. 13, no. 2, 2021.

\end{thebibliography}

\end{document}